\documentclass[aps,showkeys,floatfix]{revtex4}
\usepackage{amsmath}
\usepackage{latexsym}
\usepackage{float}
\usepackage{amssymb}
\usepackage{graphicx}
\usepackage{epsfig}
\newcommand{\tbf}{\textbf}

\begin{document}

\title{Absorption/Expulsion of Oligomers and  Linear Macromolecules in a Polymer
Brush}
\author{ A. Milchev\footnote{Corresponding author email:
milchev@ipc.bas.bg}$^{1,3}$, S. A. Egorov$^2$ and K. Binder$^1$}
\affiliation{$^1$ Institut f\"ur Physik, Johannes Gutenberg-Universit\"at,
D-$55099$ Mainz, Germany, $^2$ Department of Chemistry, University of 
Virginia, Charlottesville, Virginia 22901, USA, $^3$ Institute of Physical
Chemistry,
Bulgarian Academy of
Sciences, 1113 Sofia, Bulgaria}

\begin{abstract}

The absorption of free linear chains in a polymer brush was studied with respect
to chain size $L$ and compatibility $\chi$ with the brush by means of Monte
Carlo (MC) simulations and Density Functional Theory (DFT) / Self-Consistent
Field Theory (SCFT) at both moderate, $\sigma_g = 0.25$, and high,
$\sigma_g = 1.00$, grafting densities using a bead-spring model. Different
concentrations of the free chains $0.0625 \le \phi_o \le 0.375$ are examined.

Contrary to the case of $\chi = 0$ when all species are almost completely
ejected by the polymer brush irrespective of their length $L$, for $\chi < 0$ we
find that the degree of absorption (absorbed amount) $\Gamma(L)$ undergoes a
sharp crossover from weak to strong ($\approx 100\%$) absorption, discriminating
between oligomers, $1\le L\le 8$, and longer chains. For a moderately dense
brush, $\sigma_g = 0.25$, the longer species, $L > 8$, populate predominantly
the deep inner part of the brush whereas in a dense brush $\sigma_g = 1.00$ they
penetrate into the ``fluffy'' tail of the dense brush only. Gyration radius
$R_g$ and end-to-end distance $R_e$ of absorbed chains thereby scale with length
$L$ as free polymers in the bulk. Using both MC and DFT/SCFT methods for brushes
of different chain length $32 \le N \le 256$, we demonstrate the existence of
unique {\em critical} value of compatibility $\chi = \chi^{c}<0$. For
$\chi^{c}(\phi_o )$ the energy of free chains attains the {\em same} value,
irrespective of length $L$ whereas the  entropy of free chain displays a
pronounced minimum. At $\chi^{c}$ all density profiles of absorbing chains with
different $L$ intersect at the same distance from the grafting plane.

The penetration/expulsion kinetics of free chains into the polymer brush after
an instantaneous change in their compatibility $\chi$ displays a rather rich
behavior. We find three distinct regimes of penetration kinetics of free chains
regarding the length $L$: I ($1\le L\le 8$), II ($8 \le L \le N$), and III ($L >
N$), in which the time of absorption $\tau$ grows with $L$ at a different
rate. During the initial stages of penetration into the brush one observes a
power-law increase of $\Gamma \propto t^\alpha$ with power $\alpha \propto -\ln
\phi_o$ whereby penetration of the free chains into the brush gets {\em slower}
as their concentration rises.
\end{abstract}

\maketitle

\section{Introduction}\label{Intro}

Densely-grafted chains on nonadsorbing substrate surfaces form the so-called
``polymer brush''\cite{Alex,deGennes,Skvortsov,Cosgrove,Cates,Muthu,Murat,%
Lai,Milner,Avi,Szleifer,Klein,Grest_Book,GsG,Leger,Ruehe}. These systems find
various important applications \cite{Ruehe}, e.g. as lubricants \cite{Klein},
for colloid stabilization \cite{Napper}, for tuning of adhesion and wetting
properties \cite{Ruehe,Brown}, for improving the biocompatibility of drugs
\cite{Storm}, as protective coatings preventing protein adsorption
(``nonfouling'' surfaces) in a biological milieu \cite{Hucknall}, microfluidic
chips for biomolecule separation \cite{Wang}, etc. 

The theoretical description of the conformations of macromolecules in these
polymer brushes and their dynamics has been an active topic of research
hitherto (e.g., \cite{Kim,DID,Halperin,Egorov,Yaneva,Trombly,Dukes}; for early
reviews see \cite{Milner,Avi,Szleifer,Klein,Grest_Book,GsG,Leger}. Also the
interaction of the brushes with either the solvent molecules (e.g.
\cite{DID,Egorov}) or globular proteins \cite{Halperin} and/or other
nanoparticles (e.g., \cite{Kim,Yaneva,Currie,Chen,AM_Polymer,Matsen,Gupta}) has
found much recent attention. However, in many situations of interest there will
also occur free polymer chains in the solution, interacting with the polymers
of the brush. This interaction has received relatively less attention, apart
from the case where a polymer brush interacts with a dense polymer melt
\cite{deGennes,Gast,Chare,Witten,Zhulina,Wijmans,Aubouy,Martin,Pepin,Borukhov,%
Huang}. The latter case is particularly interesting because there is very
little interpenetration of the grafted chains in the melt and the free chains
in the brush even if their chemical nature is identical (``wetting
autophobicity'' \cite{Yerushalmi,Clarke,Reiter,Mueller}).

In contrast, scaling theory \cite{deGennes}, self-consistent field
\cite{jain09} and simulation \cite{Perahia} have predicted partial penetration
of free chains into moderately dense brushes of identical chemical nature in
semi-dilute solutions when the monomer volume fraction in solution approaches
that of the brush, and this behavior has been confirmed experimentally
\cite{Huang}. Of course, when the polymer solution is very dilute, the brush
provides a free energy barrier for penetration of free chains into it and this
limits the grafting density that can be achieved when one prepares a brush by
grafting chains from solution \cite{Ruehe} (see also some attempts to model
this process by simulations \cite{PYL,Kopf}). Similarly, since typically the
energy won by the chain end when it gets grafted is of the order of $10k_BT$
only \cite{Avi,Klein,Ruehe}, there is a nonzero probability that brush chains
get released from the grafting substrate surface and are subsequently expelled 
from the brush \cite{Wittmer}.

However, most cases studied so far refer to the situation that (apart from chain
end effects) the chains in the bulk and those in the solution are identical. It
is interesting, therefore, to consider the more general situation when the
grafted chains and those in the bulk differ in their chemical nature. Then the
problem of compatibility (traditionally modelled by introducing a Flory-Huggins
 $\chi$-parameter \cite{Flory}) between the two types of chains arises. Then,
there is also no reason to assume that the length $N$ of the grafted chains,
and the length $L$ of the free chains are equal. Such situations (in
particular, when the grafted and the free chains attract each other, $\chi <
0$) are of great interest for modern applications such as protein adsorption,
''antifouling`` surfaces \cite{Hucknall}, etc. However, to the best of our
knowledge, no systematic study of the effects of the various parameters ($N,\;
L,\; \chi$ and monomer concentration of the  free chains $\phi_o$) on the
amount of absorption and the penetration kinetics has been reported so far. The
present paper presents simulation and Density Functional Theory (DFT) results
in an effort to fill this gap. In Section \ref{Model} we describe the model and
comment on some simulation aspects; Section \ref{theory} summarizes our
theoretical approach which includes both static and dynamic versions (DDFT) of
DFT as well as Self-Consistent Field Theory (SCFT) calculations. The numerical
results are described in Section \ref{results_sec} while Section
\ref{Summary} contains a summary and discussion.

\begin{figure}[htb]
\includegraphics[scale=0.3]{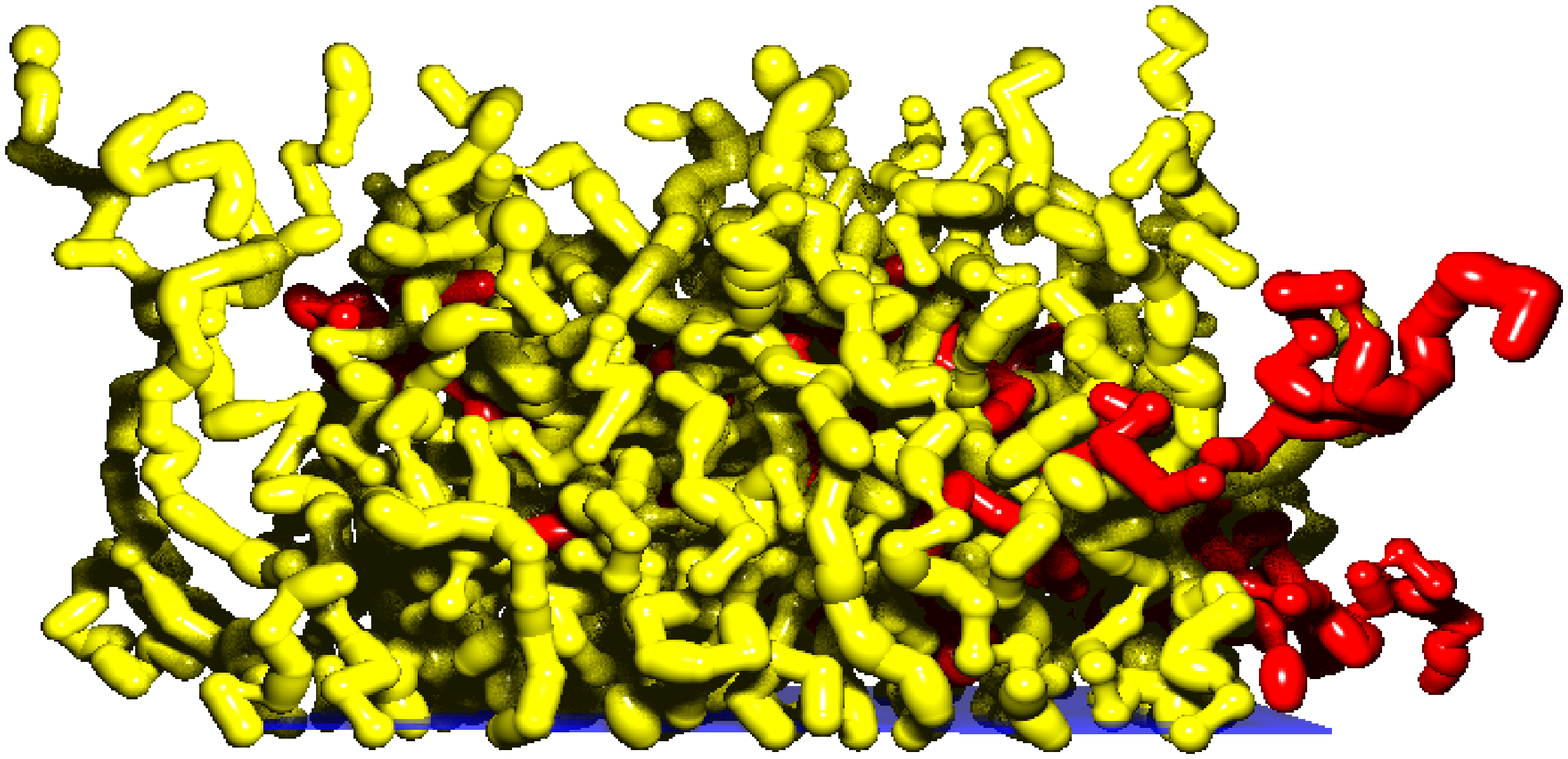}
\hspace{0.50cm}
\includegraphics[scale=0.3]{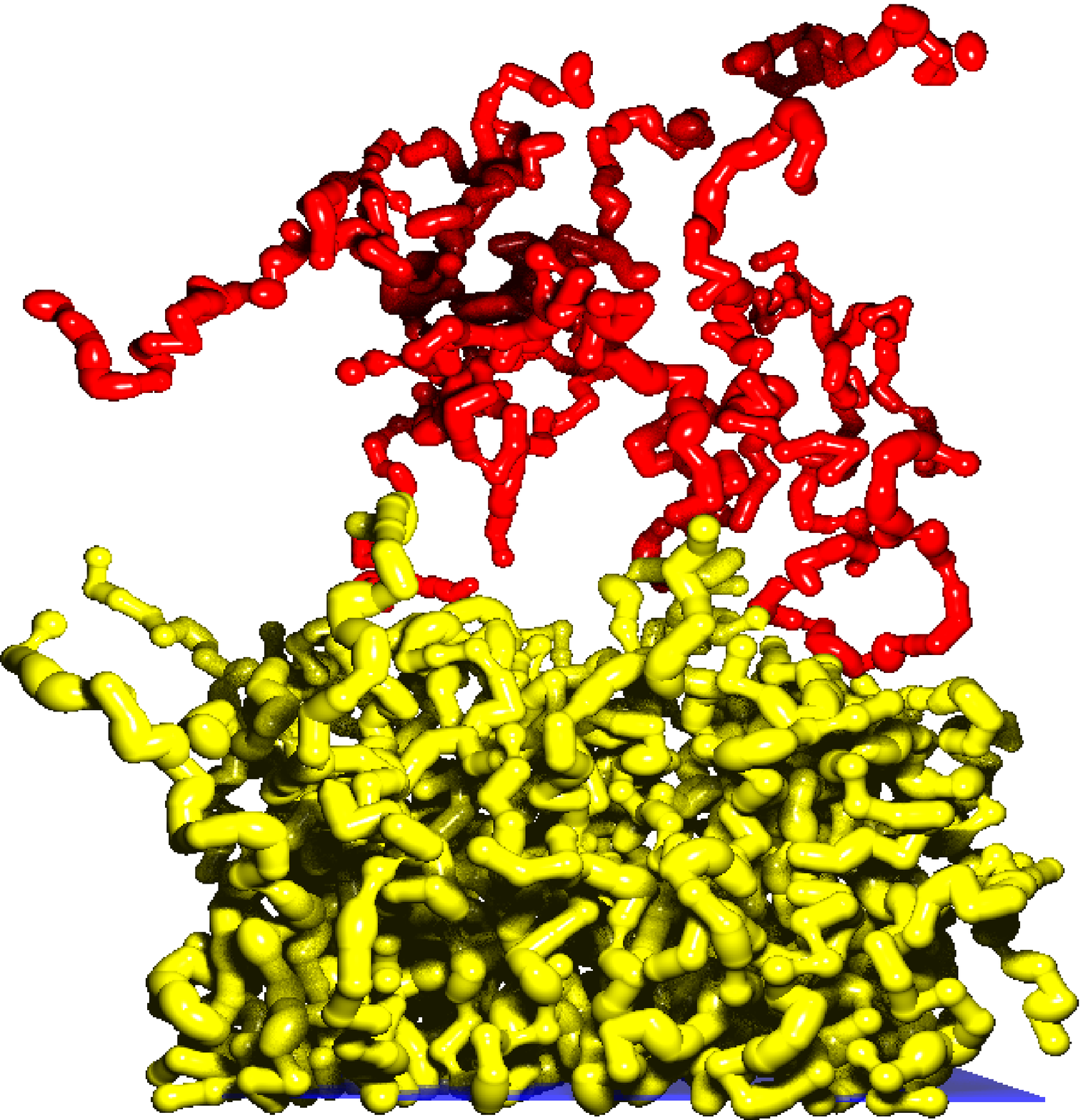}
\vspace{0.8cm} 
\caption{Snapshots of a polymer brush with chain length $N=32$ at grafting
density $\sigma_g=0.25$ and free chains of length $L=32$ at equilibrium: (left)
absorption at $\epsilon_{po}=2.00$, and (right) expulsion at
$\epsilon_{po}=0.01$. \label{snapshots_fig}}
\end{figure}

\section{Model and simulation aspects}\label{Model}

We have used a coarse grained off-lattice bead spring model\cite{AMWPKB,AMKB1}
to describe the polymer chains in our system. As far as for many applications
in a biological context rather short grafted chains are used \cite{Semal}, we
restrict ourselves to length $N=32$ of the grafted chains. The polymer brush
consists of
linear chains of length $N$ grafted at one end to a flat structureless surface.
The effective bonded interaction is described by the FENE (finitely extensible
nonlinear elastic) potential,
\begin{equation}\label{FENE}
U_{FENE}= -K(l_{max}-l_0)^2ln\left[1-\left(\frac{l-l_0}{l_{max}-l_0} \right)^2
\right]
\end{equation}
with $K=20,\; l_{max}=1,\; l_0 = 0.7,\; l_{min} = 0.4$. Thus the
equilibrium bond length between nearest neighbor monomers is $l_0 =0.7$. Here
and in what follows we use the maximal extension of the bonds,$l_{max}=1$, as
our unit length while the potential strength is measured in units of thermal
energy $k_BT$  where $k_B$ is the Boltzmann constant. 

The nonbonded interactions between brush and free chain segments are described
by the Morse potential,
\begin{equation}\label{Morse}
\frac{U_M(r)}{\epsilon_M}
=\exp[-2\alpha(r-r_{min})]-2\exp[-\alpha(r-r_{min})]\;,
\end{equation}
with $\alpha =24,\; r_{min}=0.8,\; \mbox{and} \epsilon_M/k_BT$ standing for the
strength of brush-brush, $\epsilon_{pp}$, polymer-polymer, $\epsilon_{oo}$,
and brush-polymer, $\epsilon_{po}$ interactions. In our present study we take
typically $\epsilon_{pp}=0.2$, $\epsilon_{oo}=0.1$ (that is, in the good
solvent regime with only excluded volume interactions).
For $\epsilon_{po} = 2.00, \chi = -1.85$ the free chains are absorbed in
the brush whereas for $\epsilon_{po}=0.01, \chi = -0.005$ the polymer brush
ejects them into the bulk. Note that we define here the compatibility parameter
$\chi$ simply as $\chi = 0.5(\epsilon_{pp}+\epsilon_{oo})-\epsilon_{po}$, and do
not include the coordination number (which is done when one uses lattice models
\cite{Flory}.

The size of the container is $16\times 16\times 32$. The polymer chains are
tethered to grafting sites which constitute a triangular periodic lattice on the
substrate whereby the closest distance between grafting sites is $l_{max}=1$.
Thus the largest grafting density $\sigma_g=1.0$ involves $8192$ brush segments,
if the polymer chains are anchored at distance $l_{max}=1$, and $\sigma_g =
0.25$, i.e., $2048$ brush segments, if the 'lattice constant', i.e., the
distance between adjacent head monomers on the surface is equal to $2l_{max}$.
Note that $\sigma = 1.0$ corresponds to a simulation where the monomer density
in the brush near the wall is close to the density of a polymer melt while
$\sigma = 0.25$ would correspond to a rather concentrated polymer solution.

For the chain model, $\epsilon_M/k_BT=\epsilon_{pp}=0.2$ corresponds to good
solvent conditions since the Theta-point for a (dilute) solution of polymers
described by the model, Eqs. \ref{FENE}-\ref{Morse} has been
estimated\cite{AMWPKB} as $k_B\Theta / \epsilon_M = 0.62$. In all our
simulations we use brushes formed by polymer chains consisting of $N=32$
effective monomers whereas the number of free chains $N_o$ of length $L$
(where $L$ spans the interval $1 \le L \le 64$) is taken such that the total
number of free chain segments remains constant and is equal to $512$. For a
certain length $L=64$, however, we also change the concentration of free chains
in the container by varying their number $N_o$ in the interval $8 \le N_o \le
48$. Thus, the volume fraction $c_o$ of $64$-free chains is varied between
$0.0625 \le c_o \le 0.5$. Note that, as usual, solvent molecules are not
explicitly included~\cite{AMKB1,AMKB2,AMKB3} but work which includes solvent
explicitly~\cite{DID} would  yield very similar results.

For a dense brush with polymer chains of lengths $N=32$ statistical
averages were derived from typically $10^7$ Monte Carlo Steps (MCS)
per monomer. The Monte Carlo algorithm consists of attempted moves whereby a
monomer is chosen at random and one attempts to displace it to a new randomly
chosen position $-0.5 \le \Delta x, \Delta y, \Delta z \le 0.5$ regarding the
old position. We use periodic boundary conditions in the $x-y$ directions and
impenetrable hard walls in the $z$ direction. Two typical configurations of the
polymer brush with free chains of length $L=32$, are shown in
Fig.~\ref{snapshots_fig} for the case of good, $\epsilon_{po}=2.00$, and poor,
 $\epsilon_{po}=0.01$, compatibility with the polymer brush.
 
\section{Theory}\label{theory}

We employ classical DFT to compute density profiles of free and grafted polymer
chains. Theory has been discussed in detail in previous publications, so here
we briefly summarize its most important aspects. The starting point of the DFT
treatment is the expression for the grand free energy, $\Omega$, as  a
functional of the density profiles of free and grafted chains,
$\phi_o(\tbf{R}_o)$ and $\phi_p(\tbf{R}_p)$, respectively 
($\tbf{R}_{p/o}=\{\tbf{r}_i\}$, where  
$\tbf{r}_i$ are the positions of the chain segments).
The functional $\Omega$ is related to the Helmholtz free energy
functional, $F$, via a Legendre transform:\cite{yethiraj95,woodward91}
\begin{equation}
\Omega[\phi_o(\tbf{R}_o),\phi_p(\tbf{R}_p)]=
F[\phi_o(\tbf{R}_o),\phi_p(\tbf{R}_p)]
+\sum_{\alpha=o,p}\int d\tbf{R}_{\alpha}
\phi_{\alpha}(\tbf{R}_{\alpha})V_{\alpha}(\tbf{R}_{\alpha}),
\label{omega}
\end{equation}
where $V_{\alpha}(\tbf{R}_{\alpha})$
is the external field, which in the present case is due to the hard-sphere like 
interaction of the polymer segments with the hard 
wall, $V_{p}(\tbf{R}_{p})=\sum_{i=1}^{N}v_p(z_i)$, where $v_p(z_i)=\infty$ for
$z\leq 0$ and $v_p(z_i)=0$ otherwise, with analogous expression holding for 
 $V_{o}(\tbf{R}_{o})$. Additionally, the innermost ($i=1$) bead of each
grafted chain is tethered to the wall via 
a grafting potential $\exp[-\beta v_p(z_1)/k_BT]=\delta(z_1)$,  where
$\beta=1/k_BT$. Note that the chemical
potential of both free and grafted chains is absent from the second term of 
Eq.~(\ref{omega}) because the DFT calculations are performed at a fixed number
of both free and grafted segments in order to mimic the MC simulations:
$\int_{0}^{z_{max}}dz \phi_p(z)=\sigma_g N$ and $\int_{0}^{z_{max}}dz
\phi_o(z)=N_oL/A$. in the above, $z_{max}=32$ is the box length and 
$A=256$ is the wall area. 

The Helmholtz free energy functional is separated into ideal and
excess parts,\cite{yethiraj95,woodward91} with the former given by:
\begin{equation}
\beta F_{id}[\phi_o(\tbf{R}_o),\phi_p(\tbf{R}_p)]=
\sum_{\alpha=o,p}\left\{
\int d\tbf{R}_{\alpha} \phi_{\alpha}(\tbf{R}_{\alpha})
[\ln \phi_{\alpha}(\tbf{R}_{\alpha})-1]
+\beta \int d\tbf{R}_{\alpha} \phi_{\alpha}(\tbf{R}_{\alpha}) 
V_b(\tbf{R}_{\alpha})\right\}. 
\label{fideal}
\end{equation}
where the bonding energy $V_b$ for the grafted chains is
taken as follows:
\begin{equation}
\exp[-\beta V_b(\tbf{R}_{g})]=
\prod_{i=1}^{N-1}\frac{\delta(|\tbf{r}_i-\tbf{r}_{i+1}|-b_l)}{4\pi b_{l}^{2}}, 
\label{vbond}  
\end{equation}
with a similar expression for free chains, with $N$ replaced by $L$.
This bonding potential constrains adjacent segments to a fixed separation
$b_l$. 

The excess part of the Helmholtz free energy is written as a sum of repulsive
(hard chain) and attractive terms, with the former computed in the weighted
density approximation and the latter obtained within mean-field 
approach, using Eqs.~(12)-(17) from Ref.~\cite{Egorov};  for the sake of
brevity we do not reproduce these equations here. 

The minimization of the grand free energy functional 
with respect to $\phi_p(\tbf{R}_p)$ yields
the equilibrium density distribution for the grafted chains which can
be integrated over grafting and bonding delta-functions to obtain 
the following result
for the  density profile of the $i$th segment of the 
grafted chains:\cite{Egorov} 
\begin{equation}
\phi_{pi}(z)=C_iI_p(z)I_{i}^{-}(z)I_{i}^{+}(z),
\label{phipi} 
\end{equation}
where
\begin{equation}
I_p(z)=\exp[-\beta(v_{p}(z)+\lambda_{p}(z))],
\label{ipz} 
\end{equation}
with 
\begin{equation}
\lambda_{p}(z)=\frac{\delta F_{ex}}{\delta \phi_p(z)}.
\label{lambdap}
\end{equation}
The two propagators in Eq.~(\ref{phipi}), $I_{i}^{+}$ and $I_{i}^{-}$
move from the free ($i=N$) and the tethered ($i=1$) ends of the chain,
respectively. They are computed via recursive relations given by 
Eqs.~(23)-(25) of Ref.~\cite{Egorov}. 

The normalization constant $C_i$ in Eq.~(\ref{phipi}) 
is chosen to ensure that the $i$th segment density
profile is normalized to $\sigma_g$. The total segment density profile
for the grafted chains is given by:
\begin{equation}
\phi_p(z)=\sum_{i=1}^{N}\phi_{pi}(z).
\label{phipz}
\end{equation}  
The equilibrium density profile for the segments of the free chains
can be obtained in a similar way, by minimizing 
the grand free energy functional 
with respect to $\phi_o(\tbf{R}_o)$ and integrating  over
bond-length constraining delta functions. 

The DFT equations described above are solved simultaneously to obtain the
segment density profiles for free and grafted chains. The equations are solved
iteratively using Picard algorithm,\cite{Egorov}  with the step size along the
$z$ coordinate taken to be 0.0325. The above procedure yields equilibrium
segment density profiles for a given set of interaction potentials. In addition
to the equilibrium structural properties, we have also studied the kinetics of
the adsorption of free chains into the brush, following a switch of the
interaction potential between free and grafted segments from repulsive to
attractive. To this end, we have employed the DDFT method, which is a dynamical
generalization of the DFT approach.\cite{fraaije93,xu07} MC simulations have
indicated that the segment density profiles of the grafted chains are
essentially independent of the strength of the attraction between free and
grafted segments. Accordingly, in our DDFT calculations we take $\phi_p(z)$ to
be time independent and focus on the time dependence of the free chain density,
$\phi_o(z,t)$.

The time evolution of the segment density profile of free chains is given by
the following equation:\cite{xu07}
\begin{equation}
\frac{\partial \phi_o(z,t)}{\partial \tau}=\frac{\partial}{\partial z}
\phi_o(z,t)\frac{\partial}{\partial z}\beta\mu(z,t),
\label{phipzt}
\end{equation}
where $\mu(z,t)$ is the non-equilibrium local chemical potential, and 
dimensionless time $\tau$ is defined according to 
$\tau=k_BTM/l_{max}^2t$, where $M$ is the mobility coefficient.

Initial density profile $\phi_o(z,t=0)$ corresponds to the equilibrium
distribution of free chains at a repulsive brush, i.e. $\epsilon_{po}=0.01$. At
$t=0$, the brush-free polymer attraction is instantaneously ``switched on'',
i.e. we set $\epsilon_{po}=2$. The time-dependent polymer density profile is
then propagated according to the Eq.~(\ref{phipzt}), with the time-dependent
chemical potential given by:
\begin{equation}
 \beta\mu(z,t)=\ln \phi_o(z,t)-\ln \sum_{i=1}^{L}
C_iI_o(z,t)I_{i}^{-}(z,t)I_{i}^{+}(z,t),
\label{muzt}
\end{equation}
where $I_o(z,t)$ is obtained by substituting the time-dependent density
$\phi_o(z,t)$ into the expression for $I_o(z)$ (and likewise for the
propagators $I_{i}^{-}(z,t)$ and $I_{i}^{+}(z,t)$. 
We solve Eq.~(\ref{phipzt}) using Crank-Nicholson scheme.\cite{fraaije93,xu07} 
Note that Eq.~(\ref{phipzt}) has the form of a continuity equation with the
flux (current density) given by $j(z,t)=-\phi_o(z,t)\frac{\partial}{\partial
z}\beta\mu(z,t)$. The fact that the DDFT method propagates $\phi_o(z,t)$ via
a continuity-type equation guarantees the conservation of the total number of
segments in the system, which is consistent with the simulation set-up. 

In order to compare the results of the DDFT approach with kinetic MC data, we
set the mobility coefficient $M$ equal to unity and adjust the conversion factor
between DDFT dimensionless time and kinetic MC number of steps for one
particular set of parameters $L$ and $\phi_o$. Comparisons for all other values
of $L$ and $\phi_o$ are performed using the same conversion factor, while
assuming $M$ to be inversely proportional to both $L$ and $\phi_o$. 

With the goal of shedding further light on the thermodynamic aspects of the
adsorption process, we have also performed self-consistent field theory
(SCFT) calculations of the structural properties as a function of the
interaction strength between the segments of the brush and the free chains (in
SCFT approach this interaction is characterized by the parameter $\chi$ which is
calculated in the standard fashion from the corresponding potential
well-depths: $\chi=0.5*(\epsilon_{pp}+\epsilon_{oo})-\epsilon_{po}$). 
The main motivation behind carrying out SCFT calculations is the fact that this
approach provides a more straightforward way to decompose the free energy into
entropic and energetic components, thereby providing a complementary (to
DFT) view of the adsorption process.  

The basic equations of the SCFT method are well known,\cite{Borukhov,jain09}
and will not be reproduced here for the sake of brevity. Once again, the
density profiles for free and grafted chains are written in terms of the
propagators, the only major difference from the DFT approach being that instead
of the equation of state, one employs the incompressibility constraint to set
up the equations for the density profiles, which are once again solved
iteratively using Picard's method. For example, the equation for the density
profile of the grafted chain segments takes the form:
\begin{equation}
 \phi_{p}(z)=C_{p}\sum_{i=1}^{N}\frac{I_{i}^{-}(z)I_{i}^{+}(z)}{G_{p}(z)},
\label{phipscf}
\end{equation}
 where the normalization constant $C_{p}$ is obtained from the grafting density
$\sigma_g$, and $G_{p}(z)=\exp(-\beta u_p(z))$, with 
$\beta u_p(z)=u^{\prime}(z)+\chi_{po}\phi_{o}(z)$. The hard core
potential $u^{\prime}$ is independent of the segment type and serves as a
Lagrange multiplier enforcing the incompressibility condition, meaning that
the lattice space is completely filled and no segment overlap occurs.   
The density profile of the free chain segments is obtained in a similar way.

Once the profiles are calculated, one can easily obtain excess entropy and
energy of the free chains (relative to pure unmixed components) as
follows \cite{steels00}:
\begin{equation}
S-S^{\ast}=-k_B\int_{0}^{z_{max}}dz \phi_o(z)\left\{
\frac{\ln \phi_{o}^{b}}{L}+\ln G_{o}(z)\right\},
\label{entropy}
\end{equation}
\begin{equation}
U-U^{\ast}=-k_BT\int_{0}^{z_{max}}dz \phi_o(z)\chi_{po}\phi_p(z),
\label{energy}
\end{equation}
where $\phi_{o}^{b}$ is the bulk volume fraction of free chains.

\section{Results}\label{results_sec}

\subsection{Equilibrium properties}\label{equi_ssec}

\begin{figure}[htb]
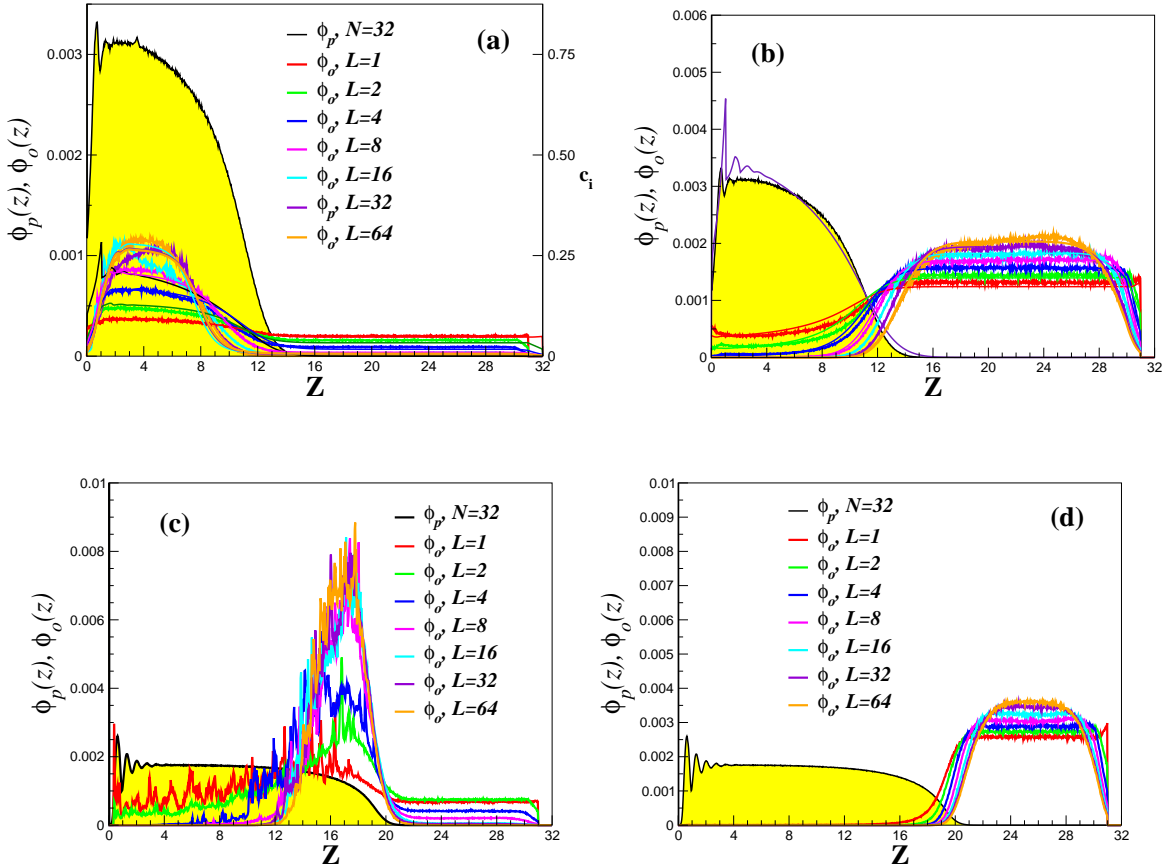

\vspace{0.7cm}
\includegraphics[scale=0.31,angle=0]{phi_int_s0.25_test.eps}
\hspace{0.30cm}
\includegraphics[scale=0.31,angle=0]{phi_rep_s0.25.eps}\\
\vspace{0.7cm}
\includegraphics[scale=0.31,angle=0]{phi_int_s1.00.eps}
\hspace{0.30cm}
\includegraphics[scale=0.31,angle=0]{phi_rep_s1.00.eps}
\vspace{0.2cm}
\caption{Density profiles of the polymer brush, $\phi_p(z)$, (shaded area) and
of free chains, $\phi_o(z)$, (thick lines) of length $L$, (given as parameter)
at two grafting densities: $\sigma_g=0.25$ (upper row), and $\sigma_g=1.00$
(lower row). (a) and (c) illustrate good compatibility between brush and
free chains, $\epsilon_{po}=2.0$ while (b) and (d) demonstrate a case of bad
compatibility, $\epsilon_{po}=0.04$. Thin solid lines in (a) and (b) denote
results from the DFT calculation. The densities in (a) are normalized so as to
reproduce the correct ratio of brush to free chains concentrations $\phi_p$
and $\phi_o$ (the absolute particle concentration $c_i$ is indicated in the
alternative $y-$axis. For the sake of better visibility, in (b), (c), and (d)
the density of all species is normalized to unit area.
\label{phi_L_fig}}
\end{figure}

In Fig. \ref{phi_L_fig} we show the density profiles of the free chains,
$\phi_o(z)$, of length $L$ for an attractive, $\epsilon_{po}=2.00$, and a
neutral, $\epsilon_{po}=0.04$, brush along with the monomer density profile of
the brush itself, $\phi_p(z)$.
Our MC simulation results indicate that at fixed segment concentration, $c_i$,
the brush profile, $\phi_p(z)$, is virtually insensitive to $L$, whereupon we
keep only one such profile in the graphs. The most striking feature which may be
concluded from Fig. \ref{phi_L_fig} is, somewhat counter-intuitively,  the
strong increase of absorption with growing length of the absorbed free chains
$L$. Evidently, both at moderate, $\sigma_g=0.25$, and high, $\sigma_g=1.00$,
grafting density, the longer polymers are entirely placed
\begin{figure}[htb]
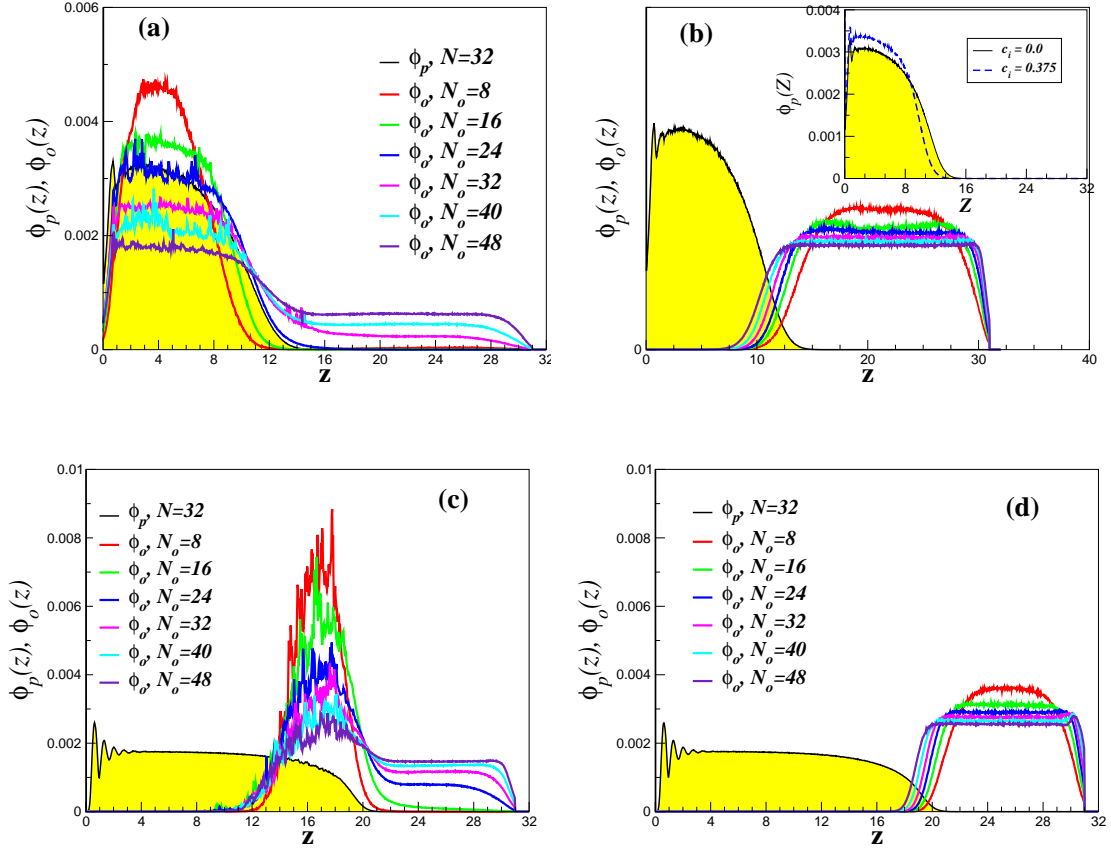

\vspace{0.7cm}
\includegraphics[scale=0.31,angle=0]{L64xNobst_s0.25_att.eps}
\hspace{0.30cm}
\includegraphics[scale=0.31,angle=0]{L64xNobst_s0.25_rep.eps}\\
\vspace{0.7cm}
\includegraphics[scale=0.31,angle=0]{L64xNobst_s1.00_att.eps}
\hspace{0.30cm}
\includegraphics[scale=0.31,angle=0]{L64xNobst_s1.00_rep.eps}
\vspace{0.5cm}
\caption{Density profiles of the polymer brush, $\phi_p(z)$, (shaded area) and
of free chains, $\phi_o(z)$, (thick lines) of length $L=64$ for different free
chain concentration (number of free chains $N_c$) (given as parameter) at two
grafting densities: $\sigma_g=0.25$ (upper row), and $\sigma_g=1.00$ (lower
row). Thin lines in (a) denote DFT results. The inset in (b) shows the change in
the density profile of brush monomers for two values of free chain
concentration: $c_i = 0.00,\;\mbox{and}\;0.375$. In (a), (c) and (d) $\phi_p$
stays practically constant. \label{dens_Nobst_fig}}
\end{figure}
inside the polymer brush whereas the much more mobile short species $L < 8$
remain uniformly distributed in the bulk above the brush end. Since some of the
absorbing chains with larger $L$ get stuck inside the brush, their density
profiles could not smoothen sufficiently for the time of the simulation run.
Therefore, we observe rather large statistical fluctuations in $\phi_o(z)$. For
repulsive brushes all species are largely expelled from the brush whereby the
situation is reversed as far as the free chain length $L$ is concerned. In the
very dense brush $\sigma_g = 1.0$, the brush profile displays the typical
oscillations near the grafting surface suggesting some layering immediately in
the vicinity of the grafting wall - Fig. \ref{phi_L_fig}c,d. In all graphs one
observes pronounced depletion effects at the upper container wall, opposing the
brush. However, the inhomogeneity of $\phi_o(z)$ near the wall at $z = 32$ has
no effect on $\phi_o(z)$ in the region of the polymer brush, the flat part of
$\phi_o(z)$ in between the brush and the confining wall at $z = 32$ is broad
enough to eliminate any finite-size effects associated with the finite linear
dimension of the simulation box in $z-$direction.

One should note also the good agreement between simulation and DFT results. In
fact, the thin lines, indicating the latter, may hardly be distinguished from
the Monte Carlo data (thick lines) in Fig. \ref{phi_L_fig}a,b. The only
significant discrepancy between theory and simulation is observed in the brush
profile in the vicinity of the grafting wall, where DFT approach overestimates
the oscillations. This discrepancy is likely due to the fact that in the DFT
method the bond lengths are constraint via delta-functions to a constant value
of $b_l=0.75$, while in the simulations the bonds are allowed to vibrate under
FENE potential, Eq. ~(\ref{FENE}). For $L=64$, Fig.~\ref{dens_Nobst_fig} shows a
qualitatively similar behavior of the density profiles for the cases of
gradually increasing free chain concentration (indicated by the number of
free chains ${\cal N}_o$ as parameter). Expectedly, for ${\cal N}_o \ge 24$
(which corresponds to monomer concentration $c_i = 0.1875$) and $\sigma_g=0.25$,
the free chains are present in the bulk over the brush as the brush interior is
then entirely filled. However, when the brush - free chain attraction increases
to $\epsilon_{po} = 3.00$, the MC data (not shown here) indicate complete
absorption of the free chains into the brush with virtually no free chains in
the bulk above the polymer brush even at the highest concentration of $c_i =
0.375$.

With increasing grafting density and/or free chain concentration, the agreement
between DFT and MC deteriorates somewhat, with the theory underestimating the
degree of penetration of free chains into the brush (see discussion of
Fig.~\ref{ads_amount_fig} below), which is likely due to the simple Tarazona's
weighting function employed in our DFT calculations. It is well known that at
higher densities it would be more appropriate to use weighting functions from
the Fundamental Measure Theory.\cite{roth10}  Indeed, precisely such approach
has been recently used to study adsorption and retention of spherical particles
in polymer brushes \cite{Borowko}.

\begin{figure}[htb]
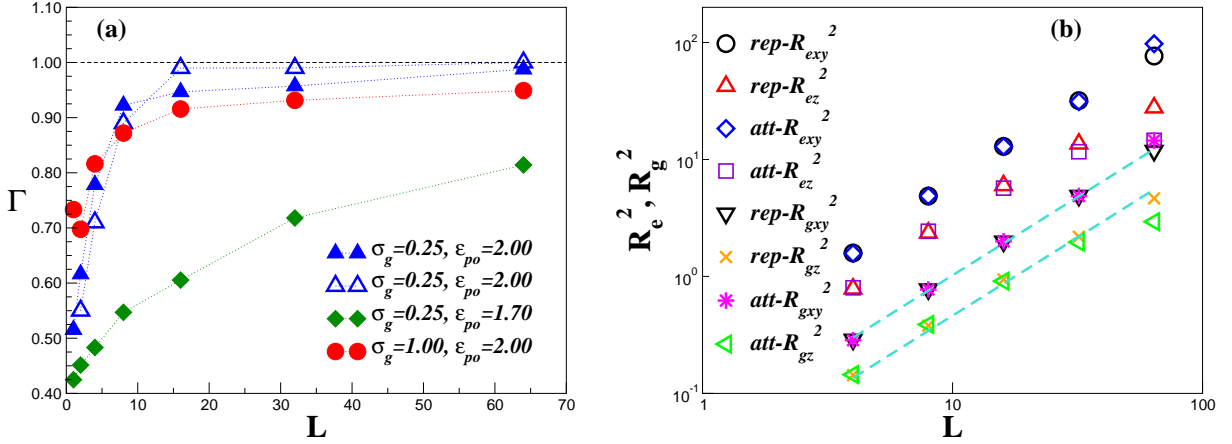

\includegraphics[scale=0.34, angle=0]{amount_L.eps}
\hspace{0.30cm}
\includegraphics[scale=0.34, angle=0]{obst_Rg_N_s0.25.eps}
\caption{(a) Variation of the absorbed amount $\Gamma$ with polymerization index
$L$ of the free chains for two grafting densities. Empty symbols denote DFT
results. The case $\epsilon_{po}=1.70$ refers to the {\em critical} degree
of brush-polymer compatibility (cf. Section \ref{crit_sect}). (b) Mean squared
radius of gyration, $R_{gxy}^2$, and end-to-end distance, $R_{exy}^2$, parallel
and perpendicular, $R_{gz}^2, \;R_{ez}^2$ to the grafting plane against length
of the free chains $L$ at grafting density $\sigma_g=0.25$. Dashed lines denote
the observed slope $\approx 1.28 \pm 0.03$. Only for the longest free chains
with $L=64$ a marked deviation from the standard scaling behavior may be
detected. \label{Rg_fig}}
\end{figure}

Next, we present MC and DFT results for the absorbed amount of free chains as a
function of degree of polymerization and  concentration. The absolute
absorbed amount is defined as the number of polymer segments located ``inside
the brush'', namely, in the region $z<z_{cut}$, where the cutoff distance
$z_{cut}$ is defined in such a way that 99\% of the brush segments are located
in the region $z<z_{cut}$. The {\em relative} absorbed amount $\Gamma$ is
defined as the ratio of the absolute absorbed amount to the total number of
free chain segments. In Fig.~\ref{Rg_fig}a one may observe the steep increase in
$\Gamma$ with growing polymer length $L$ both for brushes with $\sigma_g =
0.25$ and $\sigma_g = 1.00$ when $\epsilon_{po}=2.0$. Indeed, as indicated also
in Fig. ~\ref{phi_L_fig}, as soon as $L \ge 8$, the adsorbed amount saturates at
nearly $90\%$. A much more gradual growth of $\Gamma$ is found for the {\em
critical} attraction $\epsilon_{po}=1.70$ (see below). In Fig.~\ref{Rg_fig}a one
sees again that DFT results for the absorbed amount of polymers as a function
of the absorbate polymerization index (shown here for the case of lower grafting
density) are in good agreement with MC data, with the exception of the
intermediate-length chains, where DFT overestimates the adsorbed amount
somewhat.

\begin{figure}[htb]
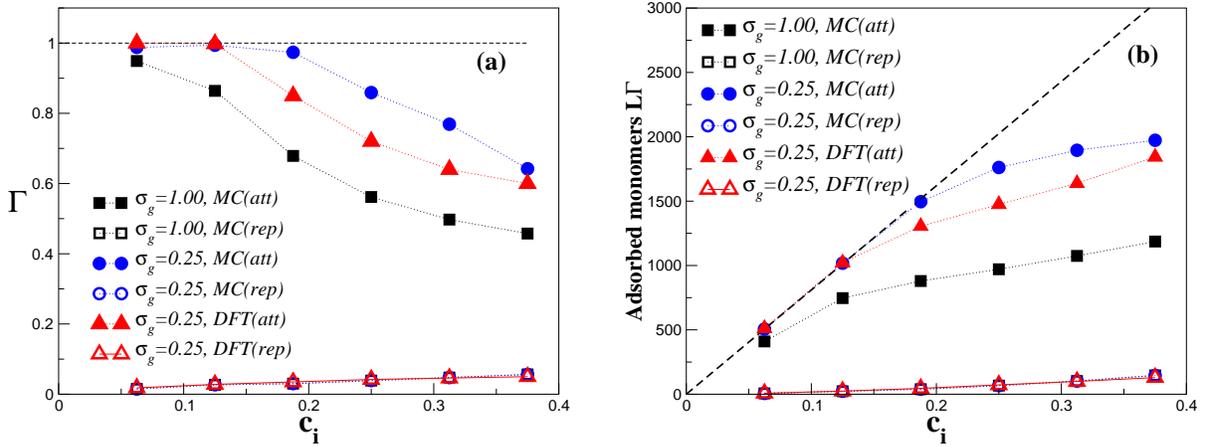

\vspace{0.8cm}
\includegraphics[scale=0.34,angle=0]{frac_phi.eps}
\hspace{0.50cm}
\includegraphics[scale=0.34,angle=0]{total_phi.eps}
\vspace{0.8cm}
\caption{Variation of the absorbed amount $\Gamma$ with free chain concentration
$c_i$ for $L=64$ and two grafting densities: $\sigma_g=0.25$ (circles), and
$\sigma_g=1.00$ (squares). Full symbols correspond to polymer absorption with
$\epsilon_{po}=2.0$ and empty symbols denote expulsion $\epsilon_{po}=0.01$. (a)
Absorbed fraction vs. $c_i$, (b) Total number of absorbed monomers against
$c_i$.
\label{ads_amount_fig}}
\end{figure}
Especially interesting is the observation, Fig.~\ref{Rg_fig}b, that the
conformations of the absorbed chains inside the brush practically do not change
with respect to those of the free chains in the bulk - the scaling behavior of
the parallel and perpendicular components of the end-to-end (squared) distance
$R_e^2$ and radius of gyration, $R_g^2$, is demonstrated in logarithmic
coordinates by straight lines whereby the value of the Flory exponent $\nu
\approx 0.64$. Due to the short lengths of the free chains used here this
value is slightly larger than what is expected for very long chains (namely
$\nu = 0.59$). Only the absorbed chains that are longer than the polymers of
the brush, $L=64 > N=32$, indicate deviations from the scaling law of single
polymers with excluded-volume interactions: the parallel component $R_{gxy}$
slightly exceeds, and the perpendicular component, $R_{gz}$, falls below the
straight line suggesting that the original shape of the $L=64$ coil flattens
parallel to the grafting plane.

Fig.~\ref{ads_amount_fig} displays the dependence of absorbed amount of
polymers on the  concentration for the highest polymerization index
studied, $L=64$.  One sees that for both grafting densities the total number of
absorbed monomers increases with concentration, while the relative absorbed
amount decreases. DFT results (again presented for the case of lower grafting
density) fall below MC data at higher concentrations, illustrating the
aforementioned observation that DFT underestimates the degree of penetration of
free chains into the brush at higher concentrations.

\begin{figure}[htb]
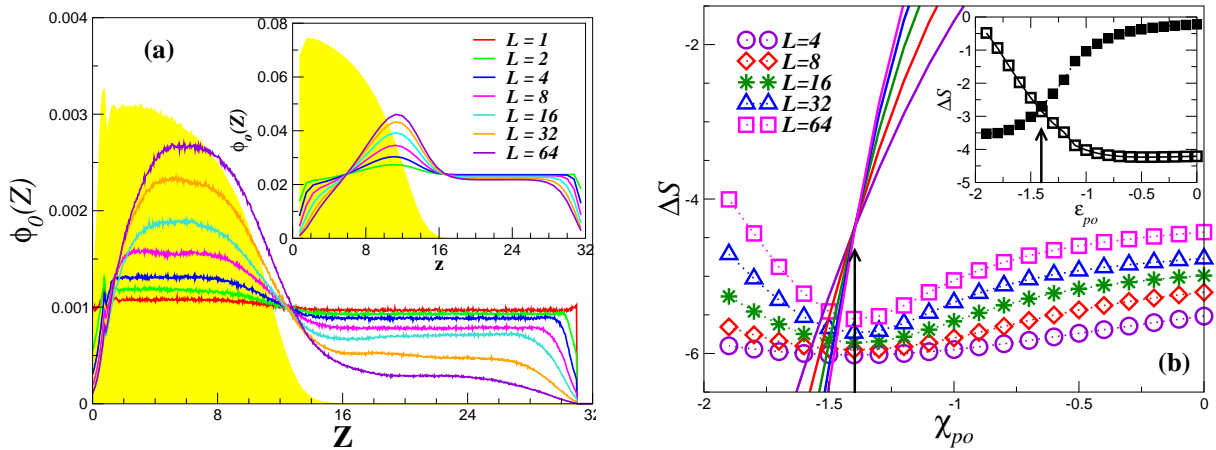

\vspace{0.8cm}
\includegraphics[scale=0.34,angle=0]{cross_L.eps}
\hspace{0.50cm}
\includegraphics[scale=0.34,angle=0]{entropy_L.eps}
\caption{(a) MC Density profiles $\phi_o(z)$ at the ``critical'' strength of
attraction $\epsilon^c_{po}=1.70$ for different lengths $L$.  (b) SCFT results
for the variation of energy (solid lines) and entropy $\Delta S$ (symbols) of
free chains of length $L$ with changing attraction $\chi_{po}$ to the
polymer brush. Arrow indicates the intersection point of energy,
$\chi^{c}_{po} =-1.40$, which coincides with the position of the minima in
$\Delta S$. All energy values are multiplied by $10$ for better visibility. In
the inset the entropy $\Delta S$ for chains with $L=64$ in the brush (full
squares) and in the bulk (empty squares) is displayed against $\epsilon_{po}$.
\label{crit_fig}}
\end{figure}

\subsection{The critical compatibility $\chi^{c}$}
\label{crit_sect}

As a remarkable feature of polymer absorption in a brush we find the existence
of a {\em critical} degree of compatibility $\chi^{c}$ between the grafted and
free chains. Fig.~\ref{crit_fig}a displays brush and free chain density profiles
for various polymer chain lengths at the critical value of the brush-polymer
attraction strength ($\epsilon^c_{po}=1.70$ for MC simulations and
$\chi^{c}_{po}=-1.40$ for SCFT calculations). While simulation and theoretical
results differ quantitatively, there is a striking qualitative similarity in
that the density profiles, irrespective of the length $L$ of the free chains,
all intersect in two single points (inside and outside the brush). The DFT
approach produces exactly the same behavior albeit for a smaller
$\epsilon^c_{po}=1.0$ (not shown here). Fig.~\ref{crit_fig}b shows SCFT results
for the excess entropy and for the internal energy per monomer (given by
Eqs.~(\ref{entropy}) and (\ref{energy}), respectively) as a function of
$\chi_{po}$. One notes immediately that all the energy curves intersect in a
single point, corresponding to $\chi_{po}=\chi^{c}_{po}$, while all the entropy
curves pass through a minimum at this point. Furthermore, the entropic curves
corresponding to the polymer segments located ``inside'' and ``outside'' the
brush (as defined earlier) intersect at the same value of $\chi_{po}$ as shown
in the inset of Fig.~\ref{crit_fig}b.

While at $\chi^{c}_{po}$ there exists thus a distance $z$ from the grafting
plane where the local concentration of polymer solutions is independent of
polymer length $L$, provided $\phi_{po}$ is kept constant for all $L$, the value
of $\chi^{c}_{po}$ itself is expected to depend on the concentration and/or the
size of the grafted chains $N$. We performed SCFT calculations to see how the
``critical'' value of $\chi^{c}_{po}$ changes with $c_i$ and $N$ within a broad
range: $0 \le c_i \le 0.375$ and $32 \le N \le 256$. We find that it increases
as $\chi^c_{po} = 1.306 + 1.326c_i + 2.393c_i^2$ with increasing free chain
concentration $c_i$, and decreases as $\chi^c_{po} = 1.874N^{-0.0858}$ with
increasing length $N$ of the grafted chains (in the latter case, the grafting
density is adjusted such that the  typical scaling variable for grafted polymers
$N\sigma_{g}^{1/3}$ is kept constant).

\subsection{Adsorption/Desorption Kinetics}
\label{kin_ssec}

Here we present our simulation and theoretical results for the 
kinetics of polymer adsorption/desorption into, or out of the brush.
\begin{figure}[htb]
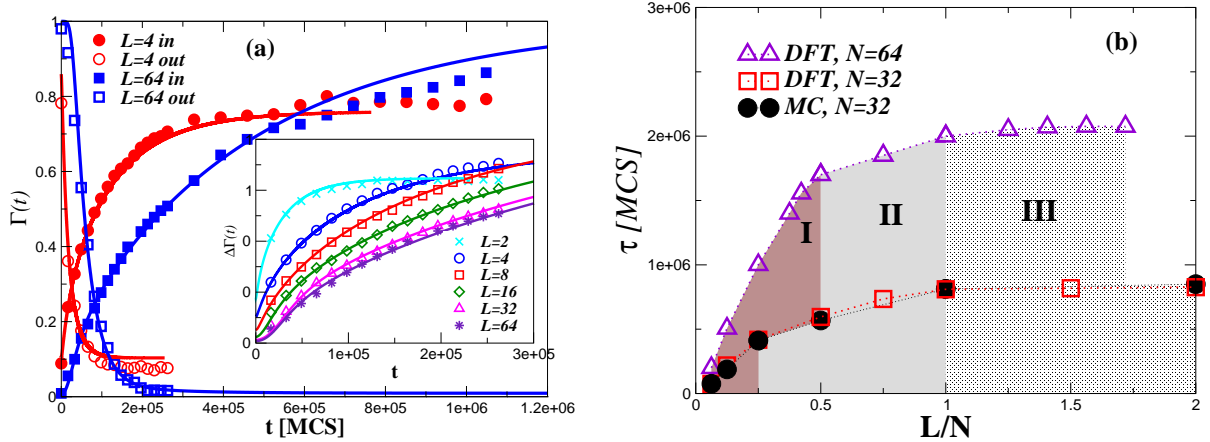

\vspace{0.8cm}
\includegraphics[scale=0.34,angle=0]{tau_in_out.eps}
\hspace{0.30cm}
\includegraphics[scale=0.34,angle=0]{tau_L.eps}
\caption{(a) Variation of the absorbed amount $\Gamma$ with elapsed time
$t$ after an instantaneous change of the interaction between brush and
free chains. Here $\sigma_g=0.25$ and the averaging was performed over $50$
cycles.  The inset shows the filling kinetics for different size $L$ of
free chains. The total number of free chain monomers $512$ was kept constant.
(b) Absorption time $\tau$ against polymer length $L$ for $\sigma_g=0.25$
displays three distinct regimes I ($1\le L\le 8$), II ($8 \le L \le N$, and III
($L > N$) (shaded areas).
\label{kin_fig}}
\end{figure}
Fig.~\ref{kin_fig}a shows the variation of the absorbed relative amount,
$\Gamma$, with elapsed time $t$ following an instantaneous switch of
the interaction between brush and free chains. As expected, the expulsion of the
adsorbate from the brush after an instantaneous switching off of brush -
polymer attraction proceeds much faster than the absorption kinetics. The
latter, as is visible from the inset to Fig.~\ref{kin_fig}a, proceeds through an
initial steep increase toward a saturation plateau of $\Gamma$ whereby the small
species absorb faster than those with larger $L$. From the intersection of
the tangent to the initial steep growth of $\Gamma$ and the saturation value one
may determine the characteristic time of absorption $\tau$ as function of $L$ -
Fig.~\ref{kin_fig}b. The results are presented for all values of free chain
lengths, and one sees that DDFT results are again in good agreement with kinetic
MC data. This also holds in Fig.~\ref{kin_fig}b where indeed the theory is in
good agreement with simulations for $N=32$. For the case of longer grafted
chains ($N$=64, $\sigma_{g}$=0.2), no simulations were performed and only
 \begin{figure}[htb]
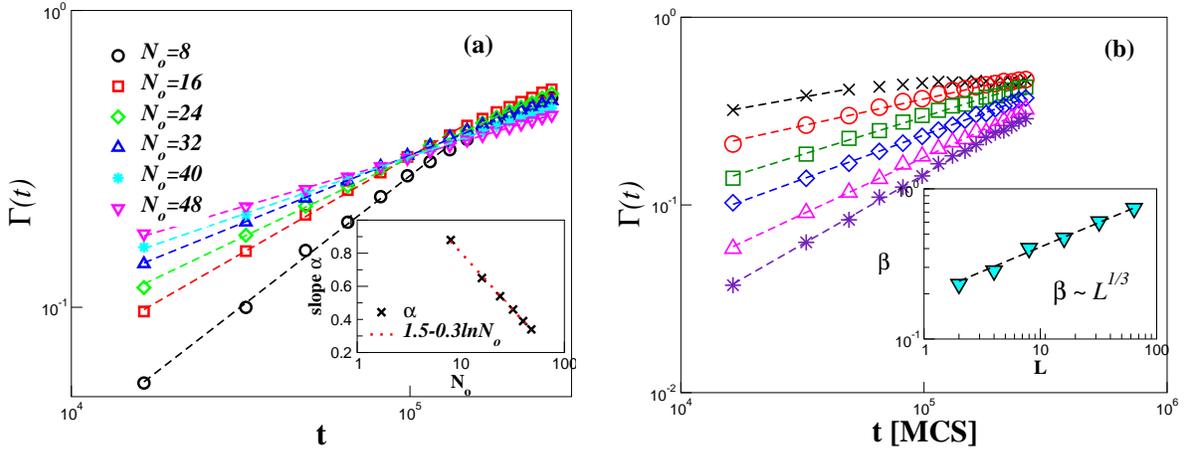

\vspace{0.8cm}
\includegraphics[scale=0.34,angle=0]{absamount_phi.eps}
\hspace{0.3cm}
\includegraphics[scale=0.34,angle=0]{Gamma_t_1.70.eps}
\caption{(a) MC data for the absorbed amount $\Gamma$ against elapsed time $t$
after the onset of absorption for different concentration of free chains with
$L=64$. The log-log plot shows that $\Gamma$ grows by power law $\Gamma(t)
\propto t^\alpha$. The measured slopes $\alpha$ are plotted in the inset against
the number of free chains $N_o$. One finds $\alpha \propto -
\frac{1}{3}\ln\phi_o$. (b) The same as in (a) but at the ``critical'' attraction
$\epsilon_{po}=1.70$ and fixed $c_i=0.0675$ where $\Gamma(t) \propto t^\beta$.
The exponent $\beta \propto L^{1/3}$ (inset).
\label{dens_fig}}
\end{figure}
theoretical predictions are shown. Nonwithstanding, for both values of $N$, one
clearly sees three regimes in the dependence of $\tau$ on $L$.  In the first
regime, the absorption time grows fast and essentially linearly with $L$ (up to
$L=8$ for the shorter brush and $L=16$ for the longer one). By analyzing the
data presented in Fig.~\ref{Rg_fig}b, one sees that this initial linear regime
corresponds to the situation when $R_g$ of absorbed chains is less than or equal
to the average distance between the grafting points. As $L$ (and, consequently,
$R_g$) is increased beyond the aforementioned values, one enters the second
regime where the growth of $\tau$, while still nearly linear, is markedly
slower. We interpret this slowing down as a halmark of an increased friction of
the penetrating coils when their radius of gyration exceeds the size of the
cavities in the polymer brush. This regime extends up to the point where the
lengths of free and grafted chains become equal. Beyond this point, for $L>N$,
the third regime is sets in, where the absorption time is essentially
independent of the free chain length. One might see therein an indication of a
change in the mechanism of free chain penetration into the brush with thickness
$H < R_g$ whereby additionally the coil flattens inside the grafted layer due to
gain in absorption energy.

Fig.~\ref{dens_fig}a displays simulation and theoretical results for the
absorption kinetics for $N=32$, $L=64$, and several values of the  concentration
$c_i = 64 N_o / 8192$. Both MC and DDFT data show that at early and intermediate
times the time dependence of the absorbed amount follows a power law $\Gamma(t)
\propto t^\alpha$. The corresponding effective exponent $\alpha$ is decreasing
as the concentration increases (see inset), although the value of $\Gamma$ at
the beginning of the intermediate time regime is larger for larger values of
$N_o$. This result is somewhat counter-intuitive, as one would expect the
driving force for absorption (and, hence, the absorption rate) to increase with
increasing concentration of free chains. A slowing down of absorption kinetics
with growing size $L$ and concentration $c_i$ of the free chains has been
experimentally observed \cite{Do} in a porous medium (activated carbon) which
resembles in certain aspects the polymer brush. In Fig.~\ref{dens_fig}b we show
the variation of the absorbed amount, $\Gamma(t) \propto t^\beta$, for the
critical attraction $\epsilon_{po}= 1.70$ -  see \ref{crit_sect}. We point out
that this well pronounced power law increase of $\Gamma$ was observed only at
this particular value of $\epsilon_{po}$ whereas for $\epsilon_{po}= 2.00$ where
most of our kinetic measurement were performed, no simple $\Gamma - t$
relationship was found - cf. Fig. \ref{kin_fig}a. Thus, in a sense, the
particular kinetics of absorption underlines the special role of the critical
compatibility between brush and free chains.

\begin{figure}[htb]
\includegraphics[scale=0.31,angle=270]{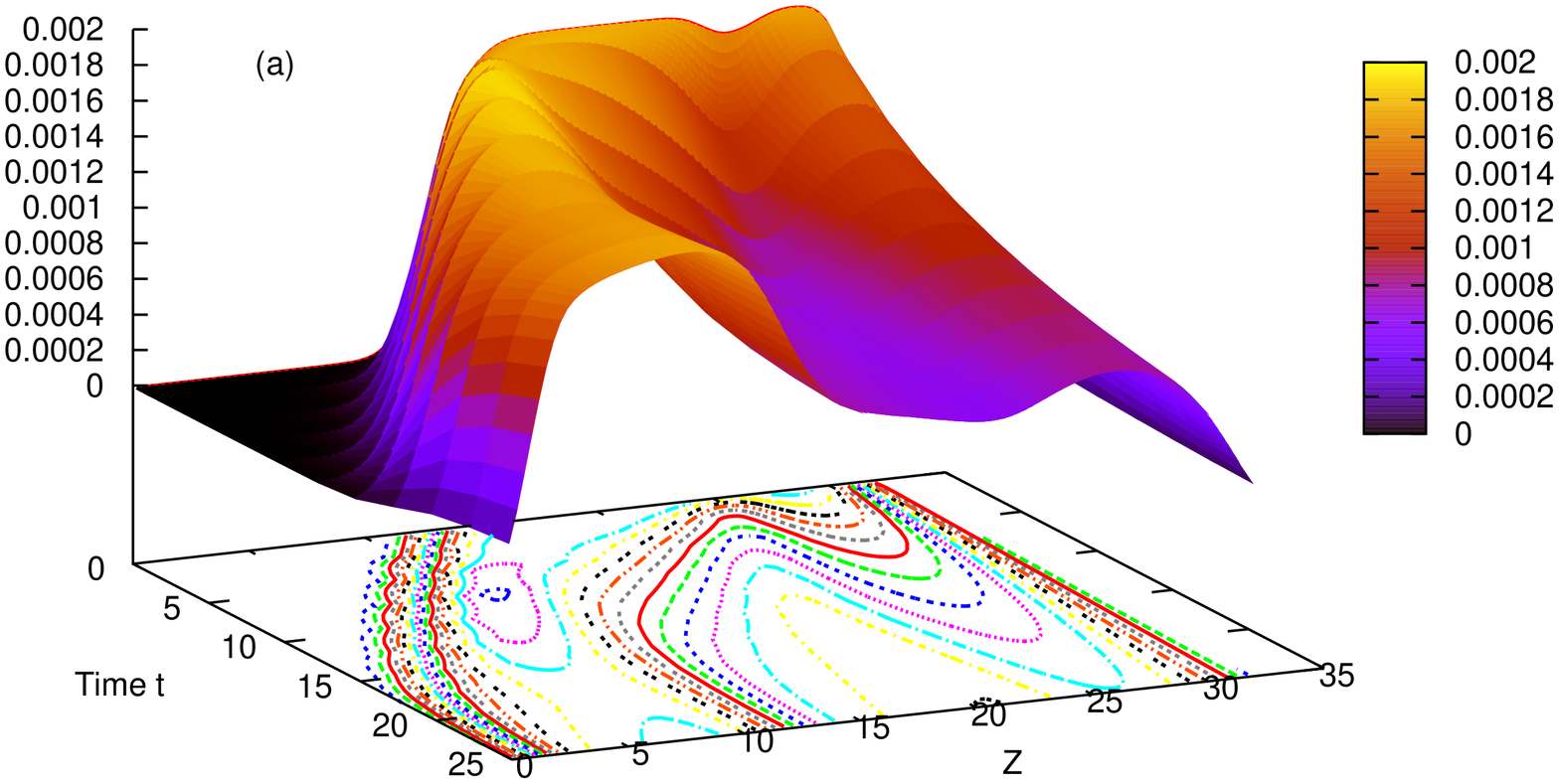}
\hspace{0.20cm}
\includegraphics[scale=0.31,angle=270]{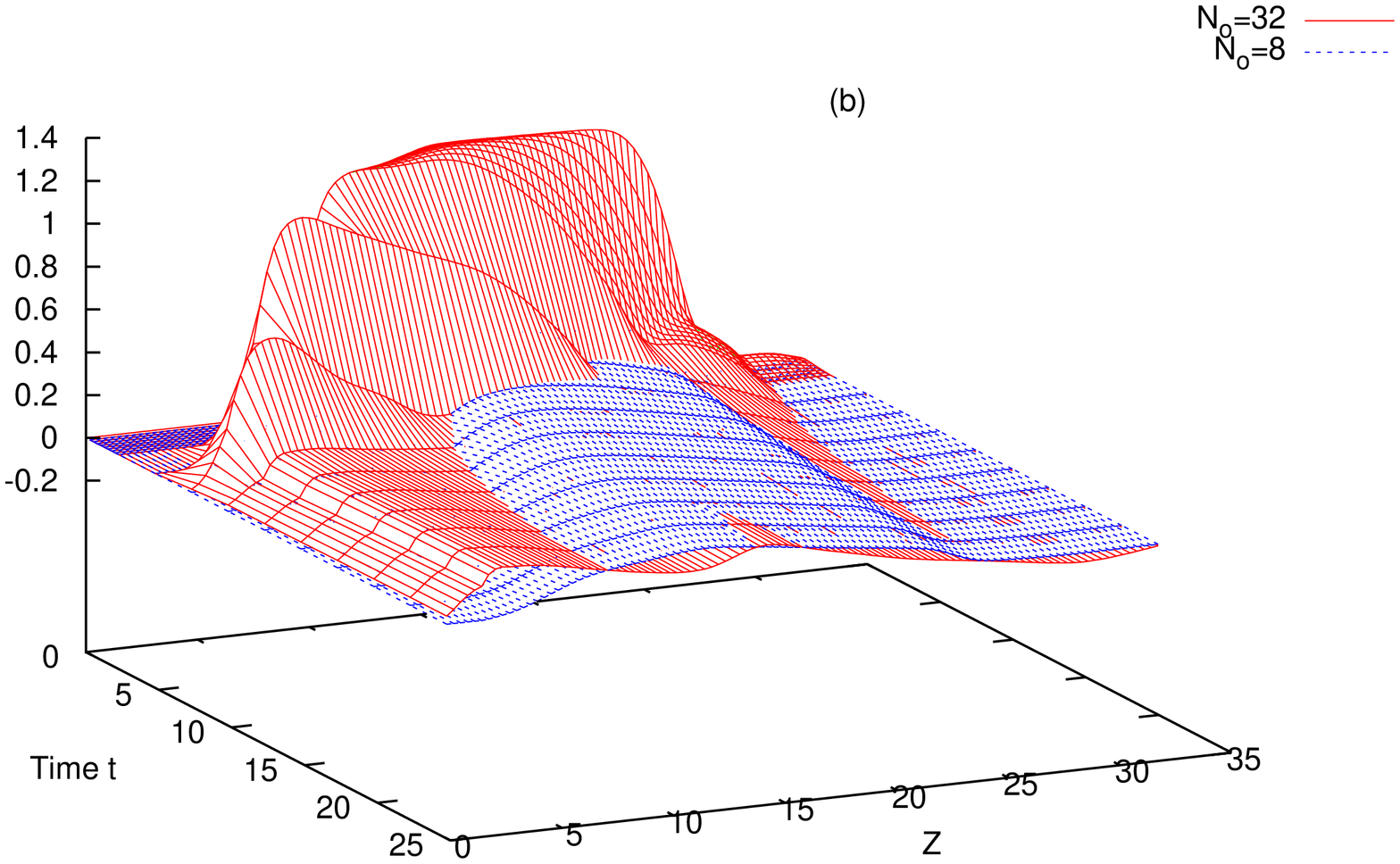}
\caption{(a) Changing concentration profile $\phi_o(z)$ of free chains with time
elapsed after a quench from $\chi=0$ to $\chi=2.0$ from DFT data. The time is
given in logarithmic coordinates. Here the mean concentration $c_o = 0.1875$ and
the time unit corresponds to 25000 MCS. The polymer brush is located at $0 \le Z
\le 12.5$. (b) Variation of the flux of free chains into the polymer brush with
time for two concentrations $c_o=0.0625,\; 0.25$ i.e., ${\cal N}_o = 8,\; 32$,
and $L=64$.
\label{kin_3d_fig}}
\end{figure}
In order to shed further light on the observed behavior, Fig.~\ref{kin_3d_fig}
shows DDFT results for the time-dependent density profile $\phi_o(z,t)$
($N_o$=24, left panel) and flux $j(z,t)$ ($N_o$=8 and 32, right panel). In the
left panel, one observes two ``ridges'' in $\phi_o(z,t)$ at all times -- a
principal ridge, initially located in the bulk above the brush, moves gradually
inside, while another (smaller) ridge is located near the opposite (bare) wall
and gradually disappears still moving in the bulk. From the right panel, one can
see that for higher concentration ($N_o$=32) the flux prevails over the lower
concentration one inside the brush and at shorter times (thereby explaining
higher initial values of $\Gamma$ seen in Fig.~\ref{dens_fig} for larger values
of $N_o$), while the situation is reversed outside the brush at longer times.
The  latter behavior is presumably due to higher mobility at lower
concentrations and explains the decrease of slope $\alpha$ with $N_o$ seen in
the inset of Fig.~\ref{dens_fig}.

\section{Discussion}\label{Summary}

In this work we studied a scarcely explored yet important aspect of oligomer
and linear macromolecule absorption in a polymer brush - the case of (more
or less) good compatibility between species in the bulk and grafted chains.
Starting from oligomers (mono- and dimers) and going up to chain lengths $L$
which exceed twice the length $N$ of the grafted chains, we have determined the
conformation of the absorbed species, the absorbed amount $\Gamma$, and
absorption kinetics (the propagation rate into the polymer brush) at different
concentration of the free chains for two cases of moderately to very dense
polymer brushes. In addition, by combining Monte Carlo simulations with DFT and
SCFT calculations, we have substantially broadened the range of lengths of the
grafted chains to $32 \le N \le 256$ in order to test more comprehensively our
findings.

The most salient, and - to some extent - unexpected features of linear chain
absorption in a polymer brush that we find are:
\begin{itemize}
 \item the dramatic increase in adsorbed amount $\Gamma(L)$ with {\em growing}
chain length $L$, and
 \item the significant slowdown of absorption kinetics with growing
concentration (i.e., with the increase of the starting gradient in density) of
the free chains
\end{itemize}
Besides these static and dynamic properties of polymer absorption in brushes, we
find that both the absorbed macromolecules and the brush itself largely retain
their structure and conformation, as seen in quantities like $R_g,\; R_e$ and
the monomer density profile $\phi_p(z)$, for different length $L$ and
concentration $\phi_o$ of the free chains, and different strength
$\epsilon_{po}$ of attraction to the grafted chains. In particular, the degree
to which the brush profile $\phi_p(z)$ is affected by absorption is found to be
much less that anticipated in some earlier theoretical predictions
\cite{Borukhov}. Nontheless, even within these small changes we observe a
slight contraction of $\phi_p(z)$ at small absorbed amounts $\Gamma$ while 
$\phi_p(z)$ gradually attains its extension roughly to that corresponding to
zero concentration of free chains with growing $\Gamma$.

An interesting finding which still needs deeper understanding is the observed
existence of a critical compatibility $\chi^c < 0$ (i.e., brush-oligomer
attraction $\epsilon_{po}^c$). At $\chi^c$ we find both in MC as well as in
DFT/SCFT that the energy of all absorbed species has a value independent of
their size $L$ whereas their entropy experiences a minimum. The critical
attraction $\epsilon_{po}^c$ is manifested by the existence of unique distance
from the grafting plane where all monomer density profiles of the free chains
intersect. Moreover, at $\epsilon_{po}^c$ the kinetics of free chain absorption
into the brush follows a clear cut power law with exponent $\beta \propto
L^{1/3}$. Undoubtedly, much more work is needed until all these fascinating new
features are fully understood.

Last not least, we emphasize the finding of three distinct regimes in the
kinetics of free chain absorption as far as the size of the free chains $L$ is
concerned. In the first regime the characteristic time for absorption $\tau$
grows rapidly with oligomer length $L$ as long as the oligomer size $R_g
\propto L^\nu \approx \sigma_g^{-1/2}$ remains smaller than the separation
between grafting sites. The second regime is marked by a slower increase of
$\tau$ with $L$ and ends roughly at $L \approx N$. The third regime of
absorption kinetics holds for $L > N$ (i.e., the penetrating free chain cannot
accommodate within the brush) and is characterized by a nearly constant $\tau$
as far as length $L$ is concerned. Interestingly, this rich kinetic behavior has
been experimentally observed in absorption in porous media \cite{Do}.

\section{Acknowledgments}
One of us, (A. M.), acknowledges support under Grant No. Bi314/22. Another, (S.
A. E.), acknowledges support from the Alexander von Humboldt foundation,
Germany.

\end{document}